\documentclass[aps,preprint,a4paper,showpacs,amsmath,amssymb,floatfix]{revtex4}

\usepackage{graphicx}
\usepackage{dcolumn}
\usepackage{bm}
\newcommand{\Op}[1]{{\boldsymbol{\mathrm{\hat{#1}}}}}
\newcommand{\beq}{\begin{equation}}
\newcommand{\eeq}{\end{equation}}
\newcommand{\beqar}{\begin{eqnarray}}
\newcommand{\eeqar}{\end{eqnarray}}
\newcommand{\bea}{\begin{eqnarray}}
\newcommand{\eea}{\end{eqnarray}}
\newcommand{\bcen}{\begin{center}}
\newcommand{\ecen}{\end{center}}

\begin{document}
\draft

\title{The minimal temperature of Quantum Refrigerators }

\author{Tova Feldmann and Ronnie Kosloff}
\affiliation{
Institute of Chemistry
the Hebrew University, Jerusalem 91904, Israel\\
}

\begin{abstract}
A first principle reciprocating quantum refrigerator is investigated to
determine the limitations of cooling to absolute zero. If the energy spectrum 
of the working medium possesses an uncontrollable gap, then there is a minimum achievable temperature above zero.
Such a gap, combined with a negligible amount of noise, prevents adiabatic following
during the expansion stage which is the necessary condition for reaching $T_c \rightarrow 0$. 
\end{abstract}
\pacs{05.70.Ln, 07.20.Pe}
\maketitle

Reciprocating refrigerators operate by a working medium shuttling heat from the cold to the hot reservoir.
This requires external control of the temperature of the  working medium. 
Upon contact with the cold side the working medium temperature  has to become 
lower than $T_c$-the cold bath temperature.  
At very low temperatures 
a quantum description of the working medium is required where the control of temperature is 
governed by manipulating the energy levels of the system.

A generic working medium possesses a Hamiltonian that is only partially controlled externally:
\begin{equation}
{\Op H }~~=~~{\Op H}_{int} + {\Op H}_{ext}(\omega)
\label{eq:genhamil}
\end{equation} 
where $\omega=\omega(t)$ is the time dependent external control field. 
Typically,  $[{\Op H}_{int},{\Op H}_{ext}] \ne 0$., therefore, $[\Op H(t), \Op H(t')] \ne 0$ as a result 
a state diagonal in the temporary energy eigenstates cannot follow adiabatically.
This fact, which is the source of quantum friction, has a profound effect on the 
performance of the heat pump \cite{k190,k201}.  
Almost perfect adiabaticity is the key to low temperature refrigeration.
Typically, the internal interaction leads to an uncontrollable  finite gap $\hbar J$
in the energy level spectrum between the ground and first excited state.  We will show that this gap 
combined with unavoidable quantum friction leads to a finite minimal temperature.
An exception is a controllable energy gap which can be reduced to zero following  the approach of the cold bath temperature $T_c$ to zero.
This case has been studied separately and leads to a vanishing rate of cooling when the absolute zero is approached \cite{k243}.

{\bf The Cycle of Operation, the Quantum Heat Pump.}
The working medium in the present study is composed of an interacting spin system.
Eq. (\ref{eq:genhamil}) is modeled by the ${\bf SU}(2)$ algebra of operators.  We can realize the model by a system of two coupled spins
${\Op H}_{int} ~~=~~\frac {1} {2} \hbar J \left({ {\boldsymbol{\mathrm{\hat
{\sigma}}}}_x^1} \otimes { {\boldsymbol{\mathrm{\hat{\sigma}}}}_x^2} -
{{\boldsymbol{\mathrm{\hat{\sigma}}}}_y^1}\otimes {\boldsymbol
{\mathrm{\hat{\sigma}}}}_y^2 ~~~
  \right) ~\equiv~\hbar J {\Op B_2}$~~~
where ${{\boldsymbol{\mathrm{\hat{\sigma}}}}}$ represents the spin-Pauli 
operators, and $J$ scales the strength of the inter particle interaction. 
For $J \rightarrow 0$, the system approaches 
a working medium with noninteracting atoms \cite{k152}.
The external Hamiltonian represents interaction of spins with an external magnetic field:
${\Op H}_{ext} ~~=~~\frac {1}{2}\hbar \omega(t)
\left({\boldsymbol{\mathrm{\hat{\sigma}}}}_z^1
\otimes {\bf \hat I^2}
+
{\bf \hat I^1} \otimes {{{{\boldsymbol{ \mathrm {  \sigma}}}}}_z^2}
\right)~\equiv~\omega(t) {\Op B_1}$. The ${\bf SU(2)}$ is closed with 
${\Op B_3}
~~=~~\frac{1}{2}  \left({ {\boldsymbol{\mathrm{\hat
{\sigma}}}}_y^1} \otimes { {\boldsymbol{\mathrm{\hat{\sigma}}}}_x^2} +
{{\boldsymbol{\mathrm{\hat{\sigma}}}}_x^1}\otimes {\boldsymbol
{\mathrm{\hat{\sigma}}}}_y^2 ~~~
  \right)$ and $[ {\Op B_1}, {\Op B_2}] \equiv  2 i {\Op B_3}$.
  
The total Hamiltonian then becomes: 
\begin{equation}
{\Op H}=\hbar \left(\omega(t) {\bf \hat B_{1}}+\rm J {\bf \hat B_{2}}\right)~~.
\label{eq:hamil}
\end{equation}
The temporary energy levels, the eigenvalues of $\Op H$ are $ \epsilon_1= - \hbar {\Omega} ,~
\epsilon_{2/3}=0,~ \epsilon_4= \hbar {\Omega} $ where $\Omega=\sqrt{\omega^2+J^2}$. For $J \ne 0$
there is a zero field splitting. Eq. (\ref{eq:hamil}) contains the essential features of the Hamiltonian
of magnetic materials \cite{oja97}.

The dynamics of the quantum thermodynamical observables are described by 
completely positive maps within the formulation of quantum open systems
\cite{lindblad76,alicki87,breuer} . The dynamics is generated by the Liouville 
superoperator, $ {\cal L}$, studied in the Heisenberg picture,
\begin{equation}
\frac {d {\Op A}}{dt}~~=~~ \frac{i}{\hbar}[{\Op H}, {\Op A}]+ {\cal L}_{D}( {\Op A})
~+~ \frac{\partial {\Op A}}{\partial t}~~~.
\label{eq:heisenberg}
\end{equation}  
where ${\cal L}_{D}$ is a generator of a completely positive Liouville super operator.

\begin{figure}[htbp]
\vspace{1cm}
\hspace{2cm}
\center{\includegraphics[height=5cm]{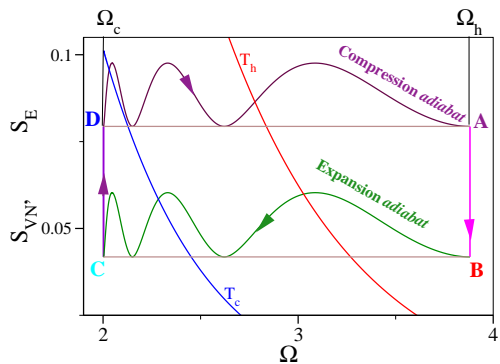}}
\caption{Refrigerator cycle in the frequency entropy plane. 
The von Neumann entropy ${\cal S}_{vn}=-tr\{ {\Op \rho}  \log {\Op \rho} \}$ ({\bf ABCD} rectangle)
as well as the energy entropy ${\cal S}_E=-\sum{ p_i \log p_i}$ are shown
($p_i$ is the population of energy level $i$). 
The hot and cold isotherms are indicated.
Both the expansion {\em adiabat} and the compression {\em adiabats} revolve exactly three periods.}
\label{fig:1}
\end{figure}

The cycle studied is composed of two segments where the working medium is 
in contact with the cold/hot baths and the external control field $\omega$ is constant, termed 
{\em isochores}. In addition, there are two segments  termed {\em adiabats} where the external field $\omega(t)$ varies and with it the energy level structure of 
the working medium. This cycle is a quantum analogue of the 
Otto cycle \cite{k176}. Each segment is characterized by a quantum propagator
${\cal U}_s$. The propagator
maps the initial state of the working medium to the final state 
on the relevant segment.
The four strokes of the cycle  (see Fig. \ref{fig:1}  ) are:

\begin{itemize}
\item{ {\em Isochore (isomagnetic)} $A \rightarrow B$: the field is maintained
constant $\omega=\omega_h$ the working medium
is in contact with the hot bath of temperature $T_h$.
${\cal L}_D$ leads to equilibrium  with heat
conductance $\Gamma_h$, for a period of $\tau_h$.  The segment dynamics is described by the
propagator ${\cal U}_h $. }
\item{ {\em Expansion adiabat (dimagnetization)} $B \rightarrow C$: The field changes
from $\omega_h$ to $\omega_c$ in a time period  of $\tau_{hc}$.
${\cal L}_D= {\cal L}_N$ represents external noise in the controls.
The propagator becomes  ${\cal U}_{hc}$ which is the main subject of study.}
\item{ {\em Isochore (isomagnetic)} $C \rightarrow D$: the field
is maintained  constant $\omega=\omega_c$ the working medium
 is in contact with the cold bath of temperature $T_c$. ${\cal L}_D$ leads to equilibrium  with heat
conductance $\Gamma_c$, for a period of $\tau_c$.  The segment dynamics is described by the
propagator ${\cal U}_c $.      }
\item{ {\em Compression adiabat (magnetization)} $D \rightarrow A$: The field  changes
from $\omega_c$ to $\omega_h$ in a time period of $\tau_{ch}$, 
${\cal L}_D= {\cal L}_N$ represents external noise in the controls.
The propagator becomes  ${\cal U}_{ch}$.}
\end{itemize}

The product of the four propagators, ${\cal U}_{s}$  is the cycle propagator: 
\begin{equation}
{\cal U}_{cyc}~~=~~{\cal U}_{ch} {\cal U}_c {\cal U}_{hc} {\cal U}_h ~. 
\label{eq:globop}
\end{equation} 
Eventually, independent of initial condition, after a few cycles, the working medium 
will reach a limit cycle characterized as  
an invariant eigenvector of ${\cal U}_{cyc} $ with eigenvalue $\bf 1$(one) \cite{k201}. 
The characteristics of the refrigerator are therefore extracted from the limit cycle.

{\em The dynamics of the expansion adiabat} .
The key to low temperatures is the expansion {\em adiabat}.  
In magnetic salt based refrigerators this segment is termed the adiabatic demagnetization stage \cite{oja97,kurti82,hakonen91}. 
A necessary condition for cooling is that the energy of the working medium at contact point $\bf C$
is lower than the equilibrium energy at temperature $T_c$.
What are the starting conditions at the beginning of the expansion segment point $\bf B$?
Considering that the efficiency 
is limited by the Carnot cycle $\eta_{otto} \le \eta_{carnot}$ leads to the condition
$\frac{\Omega_c}{\Omega_h} \le \frac{T_c}{T_h}$. Now $\Omega_c \ge J$ and $\Omega_h \ge \omega_h$ therefore
using $\Omega_c(min) =J$: 
\begin{equation}
T_c \ge J \frac{T_h}{\Omega_h}~~.
\label{eq:mt}
\end{equation} 
This condition relates the hot  end frequency $\omega_h$ to the temperature $T_c$ \cite{oja97}. 
To  force $T_c$ to zero $\omega_h \rightarrow \infty$ and with it $\Omega_h$. 
Under these conditions at equilibrium all population is in the ground state and $\langle \Op H \rangle_h = - \hbar \Omega_h$.
This is the optimal starting point for the expansion {\em adiabat}.
On the cold side the necessary condition for refrigeration is that the internal 
energy of the working medium at the end of the expansion
is smaller than the equilibrium energy with the cold bath.
\begin{equation}
\langle \Op H \rangle_c ~\le~\langle \Op H \rangle_{eq} (T_c) = - \hbar \Omega_c \left(1 -  2 e^{- \frac{\hbar \Omega_c}{k_b T_c}} \right)~,
\label{eq:fff}
\end{equation}
where $\langle \Op H \rangle_{eq}(T_c)$ is approximated by the low temperature limit $ \hbar \Omega_c \gg k_B T_c$.
Such a condition is fulfilled if the populations follows adiabatically  the ground state during the expansion {\em adiabat}.   
Then $\langle \Op H \rangle_c=-\hbar \Omega_c$ and Eq. (\ref{eq:fff}) is fulfilled. 
The expansion stage requires to reduce the external field $\omega$ from a large to a very small value maintaing adiabaticity.

The orthogonal set of  time independent operators  $ {\bf \hat B_i} $, is closed to the dynamics,
and therefore they can supply a complete vector space to represent the propagators ${\cal U}_{hc}$.
A more thermodynamically oriented alternative is 
based on a time dependent set. The set includes the energy $ {\bf \hat H} $ and two other orthogonal operators:
\begin{equation}
{\bf \hat H} \rm~ ~=~\omega (t) {\bf \hat B_1 } \rm~+~J {\bf \hat B_2 }
 \rm
~~,~~
 {\bf \hat L}  \rm~~=~-J {\bf \hat B_1 } \rm~+~\omega(t) {\bf \hat B_2}
 \rm
~~,~~
 {\bf \hat C} \rm ~ ~=~  \Omega(t)    {\bf \hat B_3 } \rm ~~.
\label{defnew} 
\end{equation} 
In general the dynamics on the expansion {\em adiabat} is generated by ${\cal L} ={\cal L}_H + {\cal L}_N$ 
where ${\cal L}_H= \frac{i }{\hbar}[\Op H, \cdot ]~$ and $\Op H(t)$ the time dependent Hamiltonian   Eq. (\ref{eq:hamil}).  
The external noise generator is ${\cal L}_N$ defined later.  
For perfect adiabatic following the propagator ${\cal U}$ factorizes between $  {\Op H } $, 
and $ {\Op L }$  and $ {\Op C } $. 

The  noiseless dynamics generated only by the Hamiltonian $\Op H(t)$  is the key to adiabaticity:
\begin{eqnarray}
\frac{d}{\Omega dt} \left( \begin{array}{c}
  {\bf \hat H }  \\
 {\bf \hat L }  \\
 {\bf \hat C }  \\
\end{array} \right)(t)=
\left(
\begin{array}{ccc}
\frac {\dot \Omega} {\Omega^2}   & - \frac {J \dot \omega}{\Omega^3}   & 0  \\
 \frac {J \dot \omega}{\Omega^3} & \frac {\dot \Omega } {\Omega^2}   &-1   \\
0  & 1 & \frac {\dot \Omega} {\Omega^2}         \\
\end{array}
\right) \left(\begin{array}{c}
{\bf \hat H }  \\
{\bf \hat L} \\
{\bf \hat C}  \\
\end{array} \right)~~.
\label{eq:newanad1} 
\end{eqnarray} 
The ability of the working medium to follow the energy spectrum is defined by the adiabatic measure $ \mu =  \frac {J \dot \omega}{\Omega^3} $. If $\mu=0$ the propagator factorizes. Constant $\mu$ minimizes the non-adiabatic deviations during the expansion. 
In addition constant $\mu$ leads to a closed form solution for the propagator ${\cal U}_{hc}$ 
forcing a particular scheduling of the external field $\omega(t)$ with time:
$\omega(t) =J {f}/\sqrt{1-f^2}$ where$f$ is a linear function of time:
$f=\frac {t}{\tau_{hc}} \left(
  \frac {\omega_c}{\Omega_c} ~-~ \frac {\omega_h}{\Omega_h}   \right) ~+~
\frac {\omega_h}{\Omega_h} $.
The adiabatic parameter $\mu$ and the time allocated to the {\em adiabat} $\tau_{hc}$,
obey the reciprocal relation: $ \mu~=~ \frac{K_{hc}}{\tau_{hc}} $
where $ K_{hc}= \frac{1}{J} \left(\frac {\omega_c}{\Omega_c} ~-~ \frac {\omega_h}{\Omega_h} \right) $.

Eq. (\ref{eq:newanad1}) is integrated by defining a new time variable:  $d \theta ~=~ \Omega dt$
The final values of $\Theta_{hc} $  becomes: $ \Theta_{hc }   ~ = ~ \tau_{hc} \frac{1}{K_{hc}} \Phi_{hc}$
where:
$
\Phi_{hc}~  = ~
\left( \arcsin(
\frac{\omega_c}{\Omega_c}) -\arcsin(\frac {\omega_h}{ \Omega_h}) \right)
$
and $ 0 \ge \Phi \ge -\frac{\pi}{2}$. 

Eq. (\ref{eq:newanad1})  is solved by noticing that the diagonal is a unit matrix
multiplied by a time dependent scalar. Therefore we seek a solution of the type 
${\cal U}_{hc}= {\cal U}_1 {\cal U}_2$ where $[ {\cal U}_1, {\cal U}_2] =0$.
The integral of the diagonal part of  Eq. (\ref{eq:newanad1} )  becomes: 
\begin{equation}
{\cal U}_1 ~~=~~e^{(\int_0^{\tau_{hc}} \frac {\dot{ \Omega}}{\Omega}dt)} {\cal I}~~=~~ 
\frac {\Omega_c}{\Omega_h} {\cal I}~~,
\label{eq:soldiag} 
\end{equation} 
which can be interpreted as the scaling of the energy levels with the variation in $\Omega$.

To integrate ${\cal U}_2$ the non diagonal parts of Eq. (\ref{eq:newanad1}),  are diagonalized, 
leading to the eigenvalues $0,-i \sqrt{q}, i \sqrt{q}$, where $q= \sqrt{1 +\mu^2}$,
and the propagator:
\begin{eqnarray}
{\cal U}_2~~=~~
\left(
\begin{array}{ccc}\frac {1 + \mu^2 c}{q^2}   & -\frac{\mu s}{q}   &  \frac {\mu (1-c)}
{q^2}   \\
 \frac {\mu s} { q}       & c    &- \frac {s}{q}   \\
 \frac {\mu (1-c)}{ q^2}  & \frac {s}{q} & \frac {\mu^2+c} {q^2}  \\
\end{array}
\right) ~~,
\label{eq:calprop}
\end{eqnarray} 
where $s=sin(q \Theta)$ and $c=cos(q \Theta)$.
 
The adiabatic limit is described by $\mu \rightarrow 0$. Then  Eq.  (\ref{eq:calprop}) factorizes. 
These are the perfect adiabatic following conditions.
In general Eq. (\ref{eq:calprop}) describes a periodic motion of $\Op H$ $\Op L$ and $\Op C$.
Each period  is defined by
\begin{equation}
q \Theta  =  ~2 \pi l~~l=0, 1,2 ...
\label{eq:mfric} 
\end{equation}  
where  $l$ is the winding number. At the end of each period  ${\cal U}_2$ restores to the identity matrix. 
These are the frictionless conditions of adiabatic following. For intermediate times $\langle {\Op H} \rangle$ is always larger than the frictionless value. The amplitude of this periodic dynamics decreases when $m$ becomes smaller, Cf. ${\cal U}_2(1,1)$ in Eq. (\ref{eq:calprop}). 

The frictionless conditions define a quantization condition for 
the adiabatic parameter $\mu$:
\begin{equation}
\mu ~~=~~ \left(~\left(\frac{2 \pi l}{\Phi_{hc}} \right)^2  -1 \right)^{-\frac{1}{2}}~~.
\label{eq:mquant}
\end{equation}
Examining Eq. (\ref{eq:mquant}) we find that there is no solution for $l=0$.
The first frictionless solution $l ~\ge ~\sqrt{\frac{\Phi_{hc}}{2 \pi}}$ leads to a minimum expansion time for frictionless solutions:
\begin{equation}
\tau_{hc}(min) = K_{hc}
\sqrt{ \left(\frac{2 \pi}{\Phi_{hc}} \right)^2  -1}~~.
\label{eq:taumin}
\end{equation}
The family of all frictionless solutions leads to refrigeration cycles which obey Eq. (\ref{eq:fff}) for any $T_c > 0$.  Such frictionless refrigerators have no
minimum temperature above $T_c=0$.

{\bf The Effective Minimal Temperature }.
Any realistic refrigerator is subject to noise on the external controls. Perfect adiabaticity requires precise control of the scheduling of the external field $\omega(t)$.  Any deviation from perfect adiabatic following,  maintaining the ground state 
on the expansion {\em adiabat} will lead to a minimum temperature.
If $\langle {\Op H } \rangle _c = -\hbar \Omega_c (1-\delta)$ where $\delta$ is the deviation from perfect adiabatic following then,
from Eq. (\ref{eq:fff}) $\delta \le 2 e^{-\frac{\hbar \Omega_c}{k_b T_c}}$, leading to:
\begin{equation}
T_c ~~\ge ~~\frac{ \hbar \Omega_c}{ - k_b \log ( \delta/2)} ~~\ge ~~  \frac{ \hbar J}{ - k_b \log ( \delta/2)} ~~.
\label{eq:tmin}
\end{equation}
We will now show that even an insignificant amount of noise will lead to $T_c(min) > 0$.

First we consider a piecewise process controlling
the scheduling of $\omega$ in time. At every time interval,
$\omega$ is updated to its new value. Then random errors are expected in the duration of these time intervals   described by the Liouville operator ${\cal L}_N$. This process is  mathematically equivalent to a dephasing 
process on the expansion {\em adiabat} \cite{k215}.  
This  stochastic dynamics can be modeled by a Gaussian semigroup with the generator \cite{gorini76,breuer}:
\begin{equation}
{\cal L}_{N_p} (\Op A)~~=~~ -\frac{\gamma_p}{\hbar^2} [ {\Op H}, [ {\Op H}, {\Op A} ]]~~,
\label{eq:h1n}
\end{equation}
which is termed phase noise. The modified equations of motion on the {\em adiabats} become:
\begin{eqnarray}
\frac{d}{\Omega dt} \left( \begin{array}{c}
 {\bf \hat H }  \\
 {\bf \hat L }  \\
  {\bf \hat C }  \\
\end{array} \right)(t)=
\left(
\begin{array}{ccc}
\frac {\dot \Omega} {\Omega^2}   & - \frac {J \dot \omega}{\Omega^3}   & 0  \\
 \frac {J \dot \omega}{\Omega^3} & \frac {\dot \Omega } {\Omega^2} -{\gamma_p}{\Omega}  &-1   \\
0  & 1 & \frac {\dot \Omega} {\Omega^2} -{\gamma_p}{\Omega}        \\
\end{array}
\right) \left(\begin{array}{c}
{\bf \hat H }  \\
{\bf \hat L} \\
{\bf \hat C}  \\
\end{array} \right)~~.
\label{eq:newanad} 
\end{eqnarray} 
We seek a product form solution: ${\cal U}_{hc}~=~{\cal U}_1 {\cal U}_2 {\cal U}_3 $ where ${\cal U}_3$ the noise propagator, given by:
$\frac{d }{ \Omega dt} {\cal U}_3 (t)~~=~{\cal W}(t) {\cal U}_3(t) $,
where:
\begin{eqnarray}
{\cal W}(t)=~~ {\cal U}_2(-t) \left(
\begin{array}{ccc}
0 & 0& 0\\
0 &- {\gamma_p}{\Omega} & 0 \\
0 & 0 & - {\gamma_p}{\Omega}
\end{array} 
\right) {\cal U}_2 (t) ~~
\label{eq:wmatrix}
\end{eqnarray}
We seek an approximate solution for ${\cal U}_3$ in the limit
when $\mu \rightarrow 0$, then  ${\cal U}_2 = {\cal I}$ since this is the frictionless limit.
Expanding Eq. (\ref{eq:wmatrix}) to first order in $\mu$ leads to:
\begin{eqnarray}
\label{eq:u3}
{\cal W}(t) \approx
-\gamma_p \Omega(t) \left(
\begin{array}{ccc}
0& \mu s& -\mu (1-c)\\
\mu s&1& 0\\
-\mu(1-c)& 0&1
\end{array} \right) ~~.
\end{eqnarray}
${\cal U}_3(\tau_{hc})$ is solved in two steps.  First evaluating the propagator for one period of $\Theta$:
for which $\Omega(t)$ is almost constant, and then the global propagator
becomes the product of the one period propagators for $l$ periods:
${\cal U}_3(\tau_{hc})\approx {\cal U}_3(\Theta=2 \pi)^l $. 
The Magnus expansion to second order is employed to obtain the one period propagator ${\cal U}_3(2 \pi )$:
\begin{equation}
{\cal U}_3(\Theta=2 \pi) ~~\approx e^{ {\cal M}_1 +{\cal M}_2 + ...}
\end{equation}
where: 
${\cal M}_1=\int_0^{2 \pi} d \Theta W(\Theta) $ and
${\cal M}_2=\frac{1}{2} \int_0^{2 \pi} \int_0^{\Theta} d \Theta d\Theta' [ {\cal W}(\Theta), {\cal W} (\Theta')]  + ...)$.
The first order Magnus term leads to:
\begin{eqnarray}
{\cal U}_3 (\Theta=2 \pi)_{M_1} \approx
 \left(
\begin{array}{ccc}
1& 0& \mu (1-e^{-2 \pi \gamma_p \Omega})\\
0&e^{-2 \pi \gamma_p \Omega} & 0\\
\mu (1-e^{-2 \pi \gamma_p \Omega})& 0&e^{-2 \pi \gamma_p \Omega }
\end{array} \right) 
\end{eqnarray}
which to first order in $\mu$, $\delta$  the deviation from perfect adiabatic following is zero.
The second order Magnus approximation leads to:
\begin{eqnarray}
{\cal U}_3 (\Theta=2 \pi)_{M_2} \approx
 \left(
\begin{array}{ccc}
C& -S& 0\\
S&C& 0\\
0& 0&1
\end{array} \right) ~~,
\label{eq:sivuv}
\end{eqnarray}
where $S= \sin \alpha$ and $C=\cos\alpha$.  
$\alpha = \pi \gamma_p \Omega \mu \sqrt{9 \mu^2+4}$ and as $\mu \rightarrow 0 $,
$\alpha = 2\pi \gamma_p \Omega \mu $.
The second order propagator  ${\cal U}_3(\tau_{hc})$, for $l$ revolutions is also a rotation matrix identical to
Eq. (\ref{eq:sivuv}), with  a new  angle 
$\alpha_l =   2 \pi \gamma_p \mu \int_0^{2 \pi l} \Omega(\Theta) d \Theta = \pi \gamma_p J \ln\left[\frac{(\Omega_h+\omega_h)(\Omega_c-\omega_c)}{(\Omega_h-\omega_h)(\Omega_c +\omega_c)}\right]$.
The deviation of ${\cal U}_3$ from the identity operator defines $\delta$. Asymptotically for $\mu \rightarrow 0~$ , 
$~\delta_{min} = 1-\cos (\alpha_l )  \approx \pi^2 \gamma_p^2 J^2  \ln[\omega_h/J]$. Any time variation in $\mu$
will lead to $\delta > \delta_{min}$.

Another source of external noise is due to amplitude errors in the control  of the
frequency $\omega(t)$. These errors are modeled by a Gaussian random process described by the Lindblad term: 
${\cal L}_{N} (\Op A ) = -\gamma_a \omega^2 [ \Op B_1 , [ \Op B_1, \Op A]]$ where $\gamma_a$ characterizes the amplitude noise.
The equation of motion for the noise propagator ${\cal U}_3$ becomes:
$\frac{d }{ d \Theta} {\cal U}_3 (\Theta)={\cal W}(\Theta)~{\cal U}_3(\Theta) $ where:
\begin{eqnarray}
\label{eq:u33} 
{\cal W}(\Theta)~~=~~ -\gamma_a \frac{\omega^2}{\Omega} {\cal U}_2(-\Theta) \left(
\begin{array}{ccc}
 \frac{J^2}{\Omega^2}& \frac{J \omega}{\Omega^2}& 0\\
\frac{J \omega}{\Omega^2} &\frac{\omega^2}{\Omega^2}& 0 \\
0 & 0 & 1
\end{array} 
\right) {\cal U}_2 (\Theta) .
\end{eqnarray}
We seek an approximate for small $\mu$. 
Using a similar procedure of calculating the propagator for one period
${\cal U}_3(\Theta=2 \pi)$. The ${\cal U}_3(1,1)$ element decouples from the remaining part of the propagator. 
As a result $1 -\delta = {\cal U}_3(1,1) \approx e^{-\gamma_a \omega_h^2 \tau_{hc}\frac{J^2}{\Omega_c^2 }} $.
The smallest $\delta$ is achieved for a one period cycle, Eq. (\ref{eq:taumin}) then 
$\delta_{min} \approx 4 \gamma_a J \frac{\omega_h^2}{\Omega_h^2}$.

\begin{figure}[htbp]
\vspace{1cm}
\hspace{2cm}
\center{\includegraphics[height=8cm]{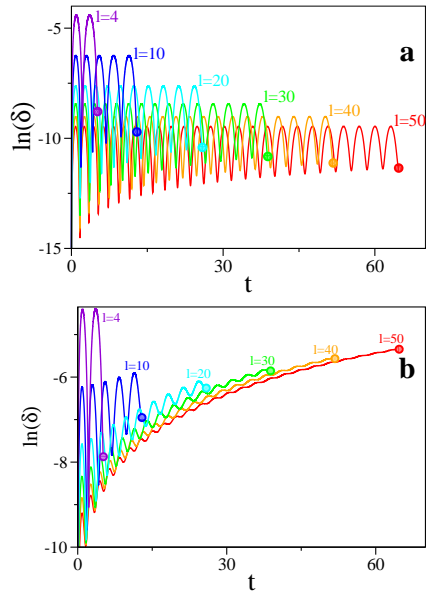}}
\caption{The deviation $\delta$ from perfect adiabatic following
as a function of the expansion time
for different revolutions $l$ for {\bf a}) phase noise and {\bf b}) amplitude noise.  The end points of each expansion are marked by a filled circle
corrosponding to the value of $\delta$ at the end of the expansion 
{\em adiabat}. For the phase noise the final $\delta$ decreses with $\tau_{hc}$
while for amplitude noise $\delta$ increases with $\tau_{hc}$.  
}
\label{fig:2}
\end{figure}

Figure \ref{fig:2} shows $\delta$ as a function of propagation time for different values of $\mu$ calculated numerically for phase (integrating Eq. (\ref{eq:u3}) ~) 
and amplitude noise (integrating Eq. (\ref{eq:u33}) ~).

\begin{figure}[htbp]
\vspace{1cm}
\hspace{2cm}
\center{\includegraphics[height=5cm]{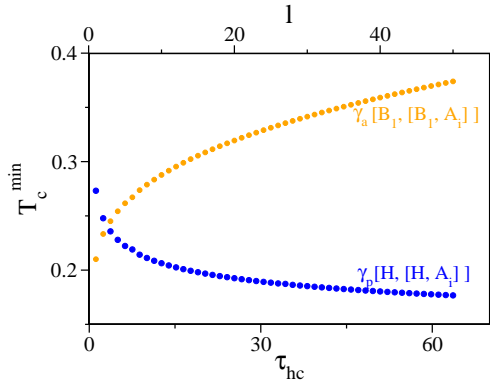}}
\caption{The minimum temperature as a function of the expansion {\em adiabat} time allocation $\tau_{hc}$ (bottom scale)
and $l$ the winding number (upper scale), for the two noise models.  
$\Omega_h= 3.8$, $\Omega_c=2.01$, $J=2$ $\gamma_p=-0.001$ and $\gamma_a=-0.0001$.}
\label{fig:3}
\end{figure}

Figure \ref{fig:3} shows the minimum temperature calculated numerically as a function of expansion time $\tau_{hc}$, for the phase and amplitude noise. The exact numerical calculation are consistent with the approximation when $\mu \rightarrow 0$.
The phase noise has a monotonic decrease of $T_c(min)$ reaching saturation as $\tau_{hc} \rightarrow \infty$
where $T_c(min)= \frac{\hbar J}{-2 k_B \log\left(\gamma_p J \Phi_{hc}/\sqrt{2} \right)}$. 
$T_c(min)$ of the amplitude noise is monotonically increasing function of time which means that short expansion times lead to the minimum temperature.
If both amplitude and phase noise operate simultaneously the minimum temperature will be obtained at the crossing point. This optimum will move with the ratio $\gamma_p/\gamma_a$.

{\bf Conclusions:} The necessary condition for the working medium to cool down to absolute zero is that  
$\langle \Op H \rangle_c = -\hbar \Omega_c$. 
In order to start in the ground state at the hot end the working medium 
has to equilibrate with a very high frequency $\Omega_h$. Perfect adiabatic following will maintain the system in its ground state
which defines the  frictionless solution. 
We found a family of additional frictionless solution obeying a quantization rule for $\mu$ the adiabatic parameter.

The main result of this study is that any noise in the controls of $\omega(t)$,
will eliminate the frictionless solutions leading to  a minimum temperature $T_c(min)$.
This finding is consistent with experiments on demagnetization cooling of a gas \cite{pfau06}
which obtained a minimum temperature an order of magnitude larger than the theoretical prediction \cite{pfau05} attributing the discrepancy to noise in the controls. 
The logarithmic dependence on the noise parameters means  that $T_c(min)$ is of the oder of $J$.
The surprise is the negative effect of phase noise. Its generator Eq. (\ref{eq:h1n}) can be interpreted as the result
of weak measurement of $\Op H$, forcing collapse to a state diagonal in energy. 
It would seem that this should improve adiabaticity and reduce friction. For an engine model we found such an effect
of enhanced performance \cite{k215}. Nevertheless even when $\mu$ is small and constant, which is the 
condition for minimal phase noise, the 
non-commutativity of $\Op H$ leads to a finite $\delta_{min}$ which saturates as 
$\tau_{hc} \rightarrow \infty$, and therefore $T_c(min) > 0$.

\section*{Aknowledgements}
We want to thank Yair Rezek and Lajos Diosi for crucial discussions.
This work is supported by the Israel Science Foundation.

\end{document}